\documentclass{article}
\usepackage{geometry}
\usepackage{fancyhdr}
\usepackage{graphicx}
\usepackage{multirow}
\usepackage{epsf}

\usepackage{latexsym}

\usepackage{amsmath}
\usepackage{amssymb}
\newcommand{\qed}{\vrule height6pt width4pt\medskip}
\newcommand{\QED}{\hfill\qed}
\newtheorem{definition}{Definition}

\newtheorem{theorem}{Theorem}

\begin{document}

\title{On the Practical Power of Automata in Pattern Matching}
\author{
\begin{tabular}{cccc}
Ora Amir\thanks{Department of Computer Science, Bar-Ilan
University, Ramat-Gan 52900, Israel,;
{\tt oramir70@gmail.com}.}
& Amihood Amir\thanks{Department of Computer Science, Bar-Ilan
University, Ramat-Gan 52900, Israel, +972 3 531-8770; {\tt
amir@cs.biu.ac.il}; and College of Computing, Georgia Tech,
Atlanta, GA 30332. Partly supported by ISF grant 1475/18 and BSF grant
2018141.}
& Aviezri Fraenkel\thanks{Dept of Computer Science and Applied Mathematics,
Weizmann Institute of Science, Rehovot 76100, Israel, +972 8 934-3545;
{\tt aviezri.fraenkel@weizmann.ac.il}.}
&
David Sarne\thanks{Department of Computer Science, Bar-Ilan
University, Ramat-Gan 52900, Israel, +972 3 531-8052; {\tt
sarned@cs.biu.ac.il}.}
\\
{\small Bar-Ilan University} & {\small Bar-Ilan University} & {\small
  Weizmann Institute of Science} & {\small Bar-Ilan University} \\
 {\small and} & {\small and} \\
 {\small NVIDIA} & {\small Georgia Tech}
\end{tabular}
}

\date{}

\maketitle

\thispagestyle{empty}

\begin{abstract}
The classical pattern matching paradigm is that of seeking
occurrences of one string - the pattern, in another - the text, where
both strings are drawn from an alphabet set $\Sigma$. Assuming the
text length is $n$ and the pattern length is $m$, this problem can
naively be solved in time $O(nm)$. In Knuth, Morris and Pratt's
seminal paper of 1977, an automaton, was developed that allows solving
this problem in time $O(n)$ for any alphabet.

This automaton, which we will refer to as the {\em KMP-automaton}, has
proven useful in solving many other problems. A notable example is the
{\em parameterized pattern matching} model. In this model, a consistent
renaming of symbols from $\Sigma$ is allowed in a match. The
parameterized matching paradigm has proven useful in problems in
software engineering, computer vision, and other applications.

It has long been suspected that for texts where the
symbols are uniformly random, the naive algorithm will perform as well
as the KMP algorithm. In this paper we examine the practical
efficiency of the KMP algorithm vs. the naive algorithm on a randomly
generated text. We analyse the time under various parameters, such as
alphabet size, pattern length, and the distribution of
pattern occurrences in the text. We do this for both the original
exact matching problem and parameterized matching. While the folklore
wisdom is vindicated by these findings for the exact matching case,
surprisingly, the KMP algorithm works significantly faster than the
naive in the parameterized matching case. 

We check this hypothesis for DNA texts, and observe a similar
behaviour as in the random text. We also show a very structured case
where the automaton is much more efficient.
\end{abstract}

\section{Introduction\label{s:introduction}}

One of the most well-known data structures in Computer science is the
Knuth-Morris-Pratt automaton, or the {KMP automaton}~\cite{KMP-77}. It
allows solving the {\em exact string matching problem} in linear
time. The exact string matching problem has input text $T$ of length
$n$ and pattern $P$ of length $m$, where the strings are composed of
symbols from a given alphabet $\Sigma$. The output is all text
locations where the pattern occurrs in the text. The naive way of
solving the exact string matching problem takes time $O(nm)$. This can
be achieved by sliding the pattern to start at every text location,
and comparing each of its elements to the corresponding text
symbol. Using the KMP automaton, this problem can be solved in time
$O(n)$. In fact, analysis of the algorithm shows that at most $2n$ 
comparisons need to be done.

It has long been known in the folklore~\footnote{The second author
heard this for the first time from Uzi Vishkin in 1985. Since then 
this belief has been mentioned, in many occasions, by various
researchers in the community.} that if the text is composed of
uniformly random alphabet symbols, the naive algorithm's time is also
linear. This belief is bolstered by the fact that the naive
algorithm's mean number of comparisons for text and pattern over a
binary alphabet is bounded by  
$$
n\sum_{i=1}^m  {i\over 2^i}\ {\rm which\ is\ bounded\ by}\ 2n\ {\rm
  comparisons}.
$$
The number of comparisons in the KMP algorithm is also bounded by
$2n$. However, because control in the naive algorithm is much simpler,
then it may be practically faster than the KMP algorithm.

The last few decades have prompted the evolution of pattern
matching from a combinatorial solution of the exact string
matching problem ~\cite{FP-74,KMP-77} to an area concerned with
approximate matching of various relationships motivated by
computational molecular biology, computer vision, and complex
searches in digitized and distributed multimedia libraries
~\cite{crochemore:rytter:book,apostolico:galil:book}.
An important type of non-exact matching is the {\em parameterized
matching} problem which was introduced by 
Baker~\cite{bak:96,bak:97}. Her main motivation lay in software
maintenance, where program fragments are to be considered
``identical'' even if variable names are different. Therefore,
strings under this model are comprised of symbols from two
disjoint sets $\Sigma$ and $\Pi$ containing {\em fixed symbols} and
{\em parameter symbols} respectively. In this paradigm, one seeks {\em
parameterized occurrences}, i.e., exact occurrences up to renaming of
the parameter symbols of the pattern string in the respective text
location. This renaming is a bijection $b: \Pi \rightarrow \Pi$. An
optimal algorithm for exact parameterized matching appeared
in~\cite{a:farach:muthu:pmatch:93}. It makes use of the KMP automaton
for a linear-time solution over fixed finite alphabet $\Sigma$.
Approximate parameterized matching was investigated
in~\cite{bak:96,hls:04,ael:07}. Idury and
Sch\"affer~\cite{Idury-Schaffer-p:93} considered multiple matching of
parameterized patterns.

Parameterized matching has proven useful in other contexts as well.
An interesting problem is searching for
color images (e.g.~\cite{swain91color,bmk:95,ACD:02}). Assume, for
example, that we are seeking a given icon in any possible color
map. If the colors were fixed, then this is exact two-dimensional
pattern matching~\cite{ABF-92}. However, if the color map is different
the exact matching algorithm would not find the pattern. Parameterized
two dimensional search is precisely what is needed. If, in addition,
one is also willing to lose resolution, then a two
dimensional function matching search should be used, where the
renaming function is not necessarily a bijection~\cite{aalp:06,an:07}.

Parameterized matching can also be naively done in time $O(nm)$. Based
on our intuition for exact matching, it is expected that here, too,
the naive algorithm is competitive with the KMP automaton-based
algorithm of~\cite{a:farach:muthu:pmatch:93} in a randomly generated
text. 

In this paper we investigate the practical efficiency of the
automaton-based algorithm vs. the naive algorithm both in exact and
parameterized matching. We consider the following parameters: pattern
length, alphabet size, and distribution of pattern occurrences in the
text.
Our findings are that, indeed, the naive algorithm is faster than the
automaton algorithm in practically all settings of the {\em exact
  matching problem}. However, it was interesting to see that the
automaton algorithm is {\em always} more effective than the naive
algorithm for {\em parameterized matching} over randomly generated
texts. We analyse the reason for this difference.

We established that the randomness of the text is what made the naive
algorithm so efficient for exact matching. We, therefore, ran the
comparison in a very structured artificial text, and the automaton
algorithm was a clear winner.  

Having understood the practical behavior of the naive vs. automaton
algorithm over randomly generated texts, we were curious if there were
``real'' texts with a similar phenomenom. We ran the same experiments
over DNA texts and observed a similar behavior as that of a randomly
generated text. 

\section{Problem Definition\label{s:def}}
We begin with basic definitions and notation generally
following~\cite{AlgorithmsOnStrings}.

Let $S=s_1 s_2\ldots s_n$ be a \textit{string} of length $|S|=n$
over an ordered alphabet $\Sigma$.  By $\varepsilon$ we denote an empty string.
For two positions $i$ and $j$ on $S$, we denote by
$S[i.. j]=s_i.. s_j$ the \textit{factor}  (sometimes called
\textit{substring}) of $S$ that begins at position
$i$ and ends at position $j$ (it equals $\varepsilon$ if $j<i$).
A {\em prefix} of $S$ is a factor that begins at
position $1$ ($S[1.. j]$) and a {\em suffix} is a factor that ends at
position $n$ ($S[i..n]$).

The {exact string matching problem} is defined as follows:

\begin{definition}\label{d:match} ({\bf Exact String Matching)}
Let $\Sigma$ be an alphabet set, $T=t_1\cdots t_n$ the {\em text} and
$P=p_1\cdots p_m$ the {\em pattern}, $t_i,p_j \in \Sigma,\ \
i=1,\ldots{},n; j=1,\ldots{},m$.
The {\em exact string matching} problem is:\\
{\bf input}: text $T$ and pattern $P$.\\
{\bf output}: All indices $j,\ \ j\in \{1,...,n-m+1\}$ such that
$$ t_{j+i-1}=p_{i},\ \ {\rm for\ } i=1,...,m $$
\end{definition}

We simplify Baker's definition of parameterized pattern
matching.

\begin{definition}\label{d:mmatch} ({\bf Parameterized-Matching})
Let $\Sigma$, $T$ and $P$ be as in Definition~\ref{d:match}.
We say that  $P$ {\em parameterize-matches}
or simply {\em $p$-matches} $T$ in location $j$ if
$p_i\cong
t_{j+i-1},\quad  i=1,\ldots{},m$, where $p_i\cong t_j$ if and only
if the following condition holds:\\
\mbox{} \qquad for every $k=1,\ldots{},i-1,\quad p_i=p_{i-k}$ if and only if
      $t_{j}=t_{j-k}$.

The {\em $p$-matching problem} is to determine all $p$-matches
of $P$ in $T$.

It two strings  $S_1$ and $S_2$ have the same length $m$ then they
are said to {\em parametrize-match} or simply {\em $p$-match}
if $s_{1_i} \cong s_{2_i}$ for all $i\in\{1,...,m\}$.
\end{definition}

Intuitively, the matching relation $\cong$
captures the notion of
one-to-one mapping between the alphabet symbols.
Specifically,
the 
condition in the definition of $\cong$
ensures that there exists a bijection between the symbols from
$\Sigma$ in the pattern and those in the
overlapping text, when they $p$-match.
The relation $\cong$ has been defined
by~\cite{a:farach:muthu:pmatch:93} in  a manner
suitable for computing the bijection.

{\bf Example:} The string $ABABCCBA$ parameterize matches the string
$XYXYZZYX$. The reason is that if we consider the bijection
$\beta:\{A,B,C\} \rightarrow \{X,Y,Z\}$ defined by 
$A\xrightarrow{\beta} X,\ \ B\xrightarrow{\beta}
Y,\ \ C\xrightarrow{\beta} Z$, then we get 
$\beta(ABABCCBA) = XYXYZZYX$. This explains the requirement in
Def.~\ref{d:mmatch}, where two sumbols match iff they also match in
all their previous occurrences.

Of course, the alphabet bijection need not be as extreme as
bijection $\beta$ above. String $ABABCCBA$ also parameterize matches
$BABACCAB$, because of bijection $\gamma:\{A,B,C\} \rightarrow \{A,B,C\}$
defined as: $A\xrightarrow{\gamma} B,\ \ B\xrightarrow{\gamma}
A,\ \ C\xrightarrow{\gamma} C$. 

\begin{figure}[htb]
\centering
\begin{minipage}{10cm}
\includegraphics[width=\textwidth]{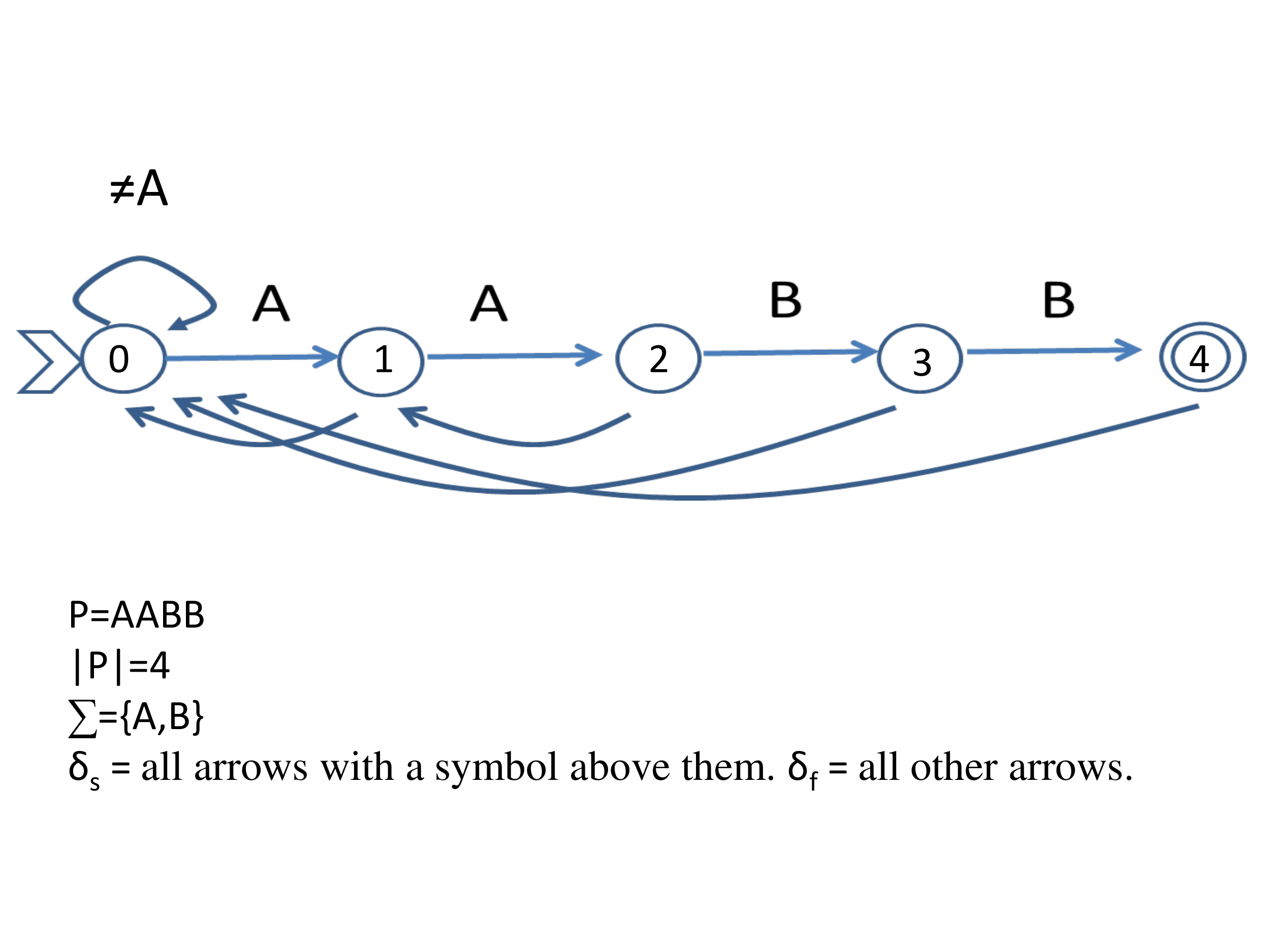}
\caption{Automaton example}
\label{f:auto}
\end{minipage}
\end{figure}

For completeness, we define the {\em KMP automaton}.
\begin{definition}\label{d:kmp}
Let $P=p_1 \ldots p_m$ be a string over alphabet $\Sigma$. The {\em KMP
  automaton} of $P$ is a 5-tuple $(Q,\Sigma,\delta_s, \delta_f, q_0,
q_a)$, where $Q=\{0,...,m\}$ is the set of {\em states}, $\Sigma$ is
the {\em alphabet}, $\delta_s:Q \rightarrow Q$ is the {\em success
  function}, $\delta_f:Q \rightarrow Q$ is the {\em failure function},
$q_0=0$ is the {\em start state} and $q_a=m$ is the {\em accepting state}.

The {\em success function} is defined as follows:\\
$\delta_s(i)=i+1$, $i=0,...,m-1$ and\\
$\delta_s(0)=0$

The {\em failure function} is defined as follows:\\
Denote by $\ell(S)$ the length of the longest proper prefix of string
$S$ (i.e. excluding the entire string $S$) which is also a suffix of
$S$.\\
$\delta_f(i)=\ell(P[1..i]),\ \ {\rm for\ } i=1,..m$.
\end{definition}
For an example of the KMP automaton see Fig.~\ref{f:auto}.
\begin{theorem}\label{t:kmp}~\cite{KMP-77}
The KMP automaton can be constructed in time $O(m)$.
\end{theorem}

\section{The Exact String Matching Problem\label{s:exact}}

The {\em Knuth-Morris-Pratt (KMP) search algorithm} uses the KMP
automaton in the following manner:

{\bf Variables:}\\
$pointer_t$ points to indices in the text.
$pointer_p$ points to indices in the pattern.

{\bf Initialization:}\\
{\sf set pointer} $pointer_t$ {\sf to} $1$.
{\sf set pointer} $pointer_p$ {\sf to} $0$.

{\bf Main Loop:}\\
{\sf While} $pointer_t \leq n-m+1$ {\sf do:}\\
\mbox{} \qquad {\sf If} $t_{pointer_t} = \delta_s(pointer_p)$ {\sf then
  do:}\\
\mbox{} \qquad \qquad $pointer_t \leftarrow pointer_t+1$\\
\mbox{} \qquad \qquad $pointer_p \leftarrow \delta_f (pointer_p)$\\
\mbox{} \qquad \qquad {\sf If} $pointer_p=m-1$ {\sf then do:} \\
\mbox{} \qquad \qquad \qquad {\sf output}
``pattern occurrence ends in text location $pointer_t$''.\\
\mbox{} \qquad \qquad \qquad $pointer_p \leftarrow \delta_f (m)$\\
\mbox{} \qquad \qquad {\sf enddo}\\
\mbox{} \qquad {\sf enddo}\\

\mbox{} \qquad {\sf else} ($t_{pointer_t} \neq \delta_s(pointer_p)$)
     {\sf do:}\\
\mbox{} \qquad \qquad {\sf if} $pointer_p = 0$ {\sf then} $pointer_t
\leftarrow pointer_t+1$\\
\mbox{} \qquad \qquad {\sf else} $pointer_p
\leftarrow \delta_f (pointer_p)$\\
\mbox{} \qquad {\sf enddo}\\
\mbox{} \qquad {\sf go to beginning of while loop}\\
\mbox{} {\sf endwhile}\\

\begin{theorem}\label{t:kmpsearch}~\cite{KMP-77}
The time for the KMP search algorithm is $O(n)$.
In fact, it does not exceed $2n$ comparisons.
\end{theorem}

\section{The Parameterized Matching Problem}\label{s:param}

Amir, Farach, and Muthukrishnan~\cite{a:farach:muthu:pmatch:93}
achieved an optimal time algorithm for parameterized string matching
by a modification of the KMP algorithm. In fact, the algorithm is
exactly the KMP algorithm, however, every equality comparison
``$x=y$'' is replaced by ``$x\cong y$'' as defined below.

{\bf Implementation of ``$x\cong y$''}

Construct table $A[1],\ldots{},A[m]$ where $A[i]=$ the largest
$k,\quad 1\leq k < i$, such that $p_k=p_i$. If no such $k$ exists then
$A[i]=i$.

The following subroutines compute ``$p_i\cong t_j$'' for $j\geq i$,
and ``$p_i\cong p_j$'' for $j\leq i$.

{\sf Compare($p_i$,$t_j$)}\\
\mbox{} \qquad {\sf if $A[i]=i$ and $t_j\neq
t_{j-1},\ldots{},t_{j-i+1}$ then return {\em equal}}\\
\mbox{} \qquad {\sf if $A[i]\neq i$ and $t_j=t_{j-i+A[i]}$ then return
{\em equal}}\\
\mbox{} \qquad {\sf return {\em not equal}}\\
{\sf end}

{\sf Compare($p_i$,$p_j$)}\\
\mbox{} \qquad {\sf if ($A[i]=i$ or $i-A[i]\geq j$) and $p_j\neq
p_1,\ldots{},p_{j-1}$ then return {\em equal}}\\
\mbox{} \qquad {\sf if $i-A[i]< j$ and $p_j=p_{j-i+A[i]}$ then return
{\em equal}}\\
\mbox{} \qquad {\sf return {\em not equal}}\\
{\sf end}

\begin{theorem}\label{t:pm}~\cite{a:farach:muthu:pmatch:93}
The $p$-matching problem can be solved in $O(n\log\sigma)$ time,
where $\sigma= {\rm min} (m,|\Sigma |)$.
\end{theorem}
\begin{proof}

The table $A$ can be constructed in
$O(m\log\sigma)$ time as follows: scan the pattern left to right
keeping track of the distinct symbols from $\Sigma$ in the pattern
in a balanced tree, along with the
last occurrence of each such symbol in the portion of the pattern
scanned thus far.
When the symbol at location $i$  is scanned, look up this symbol
in the tree for the immediately preceding occurrence; that gives
$A[i]$.

{\sf Compare} can clearly be implemented in time $O(\log \sigma)$. For
the case $A[i]\neq i$, the comparison can be done in time $O(1)$.
When scanning the text from left to right, keep the last $m$ symbols
in a balanced tree. The check
$t_j\neq t_{j-1},\ldots{},t_{j-i+1}$ in
{\sf Compare($p_i$,$t_j$)}
can be performed in $O(\log \sigma)$ time  using this information.
Similarly,
{\sf Compare($p_i$,$p_j$)}
can be performed  using $A[i]$.
Therefore,
the automaton construction in KMP algorithm with
every equality comparison ``$x=y$'' replaced by ``$x\cong y$'' takes
time $O(m\log \sigma)$ and the text
scanning takes time $O(n\log \sigma)$, giving a total of $O(n\log
\sigma)$ time.

As for the algorithm's correctness, Amir, Farach and
Muthukrishnan showed that the failure link in automaton node $i$
produces the largest prefix of $p_1\cdots p_i$ that $p$-matches the
suffix of $p_1\cdots p_i$. $\QED$
\end{proof}

\section{Our Experiments}\label{s:exp}

Our implementation was written in $C++$. The platform was Dell
latitude 7490 with intel core i7 - 8650U, 32 GB RAM, with 8 MB
cache. The running time was computed using the {\sf chrono} high
resolution clock. The random strings were generated using the {\sf
  random} Python package. 

We implemented the naive algorithm for exact string matching and for
parameterized matching. The same code was used for both, except for
the implementation of the equivalence relation for parameterized
matching, as described above. This required implementing the $A$
array. We also implemented the KMP algorithm for
exact string matching, and used the same algorithm for parameterized
matching. The only difference was the implementation of the
equivalence parameterized matching relation.

The text length {$n$} was 1,000,000 symbols. Theoretically, since both
the automaton and naive algorithm are sequential and only consider a
window of the pattern length, it would have been sufficient to run the
experiment on a text of size twice the pattern~\cite{AF-95}. However,
for the sake of measurement resolution we opted for a large text. Yet
the size of 1,000,000 comfortably fits in the cache, and thus we avoid
the caching issue. In general, any searching algorithm for {\bf
  patterns} of length less than 4MB would fit in the cache if
appropriately constructed in the manner of~\cite{AF-95}. Therefore our
decision gives as accurate a solution as possible.

We ran patterns of lengths $m=32,64,128,256,512,$ and $1024$. The
alphabet sizes tested were $|\Sigma|=2,4,6,8,10,20,40,80,160,320$. For
each size, 10 tests were run, for a total of 600 tests.

{\bf Methodology:} We generated a uniformly random text of length
$1,000,000$. If the pattern would also be randomly generated, then it
would be unlikely to appear in the text. However, when seeking a
pattern in the text, one assumes that the pattern occurs in the
text. An example would be searching for a sequence in the DNA. When
seeking a sequence, one expects to find it but just does not know
where. Additionally, we considered the common case where one does not
expect many occurrences of the pattern in the text. Consequently, we
planted $100$ occurrences of the pattern in the text at uniformly
random locations. The final text length was always $1,000,000$. The
reason for inserting 100 pattern occurrences is the following. We do
not expect manny occurrences, and a 100 occurrences in a
million-length text means that less than $0.1\%$ of the text has
pattern occurrences. On the other hand, it is sufficient to introduce
the option of actually following all elements of the pattern 100
times. This would make a difference in both algorithms. They would
both work faster if there were no occurrences at all.

We  also implemented a variation where half of the pattern occurrences 
were in the last quarter of the text. For each alphabet size and
pattern length we generated 10 tests and considered the average result
of all 10 tests. It should be noted that from a theoretical point of
view, the location of the pattern should not make a difference. We
tested the different options in order to verify that this is, indeed,
the case.

\subsection{Exact Matching}\label{ss:ext}

\subsubsection{Results} \label{sss:ext_res}

Tables~\ref{t:ext_tbl1} and ~\ref{t:ext_tbl2} in the Appendix show the
alphabet size, the pattern length, the average of the running times of
the naive algorithm for the 10 tests, the average of the running time
of the KMP algorithm for the 10 tests, and the ratio of the naive
algorithm running time over the KMP algorithm running time. Any ratio
value {\em below} $1$ means that the {\em naive algorithm is
  faster}. A {\em small} value indicates a {\em better} performance of
the naive algorithm. Any value above $1$ indicates that the KMP
algorithm is faster than the naive algorithm. The larger the number,
the better the performance.

\begin{figure}[htb]
\centering
\begin{minipage}{10cm}
  \includegraphics[width=\linewidth]{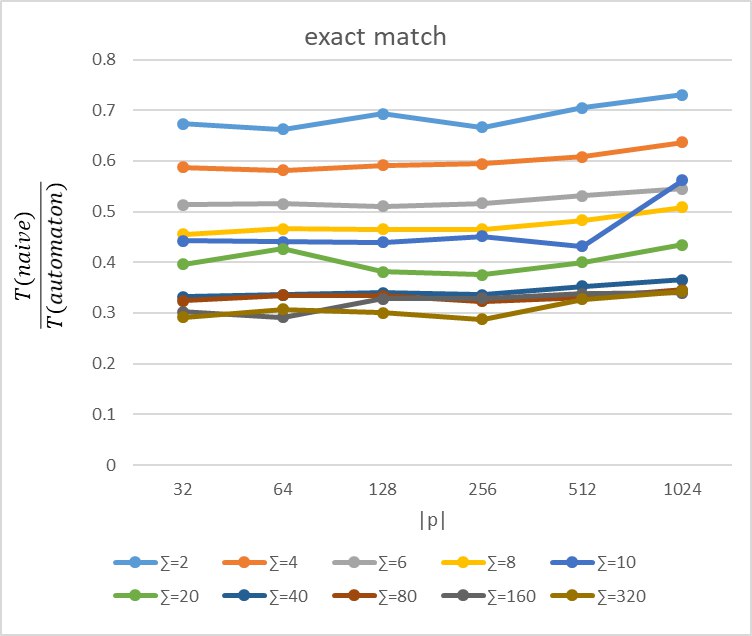}
  \caption{Performance in the Exact Matching case, pattern occurrences
    distributed uniformly random.}
  \label{f:exactu}
\end{minipage}
\end{figure}

\begin{figure}[htb]
\centering
\begin{minipage}{10cm}
  \includegraphics[width=\linewidth]{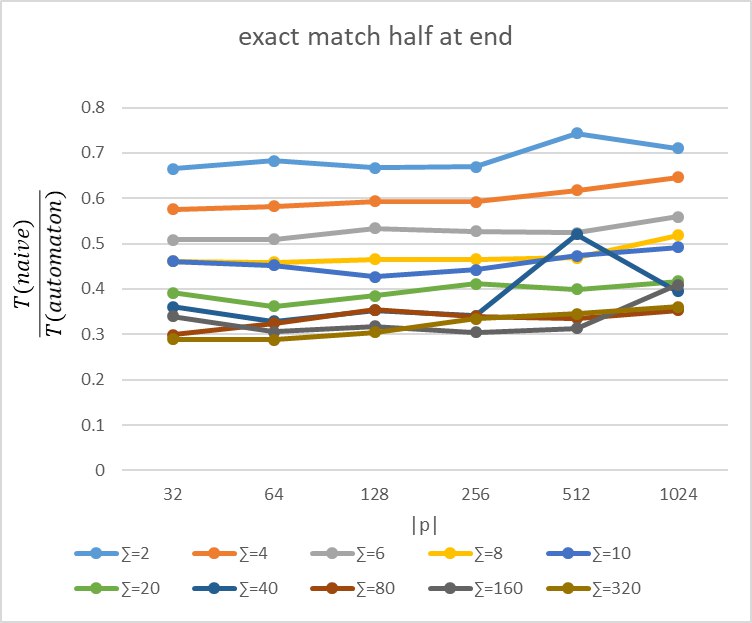}
  \caption{Performance in the Exact Matching case, pattern occurrences
  congregated at end of text.}
  \label{f:exacte}
\end{minipage}
\end{figure}

\begin{figure}[htb]
\centering
\begin{minipage}{10cm}
  \includegraphics[width=\linewidth]{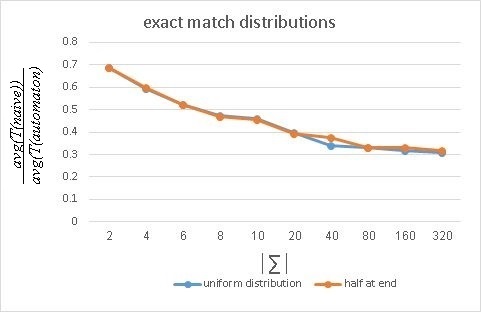}
  \caption{Comparison of average performance of uniform pattern
    distribution vs. pattern occurrences congregated at end of text.}
  \label{f:exactb}
\end{minipage}
\end{figure}

To enable a clearer understanding of the results, we present them below in
graph form. The following graphs show the results of our
tests for the different pattern lengths. In Figs.~\ref{f:exactu}
and ~\ref{f:exacte}, the $x$-axis is the pattern 
size. The $y$-axis is the ratio of the naive algorithm running time to
the KMP algorithm running time.  The different colors depict alphabet
sizes. In Fig.~\ref{f:exactu}, the patterns were inserted at random,
whereas in Fig.~\ref{f:exactb} the patterns 
appear at the last half of the text. 

To better see the effect of the pattern distribution in the
text, Fig.~\ref{f:exactb}  maps, {\em on the same graph}, both
cases. In this graph, the $x$-axis is the average running time of {\em
  all} pattern lengths per alphabet size, and the $y$-axis is the
ratio of the naive algorithm running time to the KMP algorithm running
time. The results of the uniformly random distribution are mapped in
one color, and the results of all pattern occurrences in the last half
of the text are mapped in another.

We note the following phenomena:
\begin{enumerate}
\item The naive algorithm {\bf performs better} than the
  automaton algorithm. Of the $600$ tests we ran, there were only $3$
  instances where the KMP algorithm performed better than the
  naive, and all were subsumed by the average. In the vast majority of
  cases the   naive algorithm was superior by far.
\item The naive algorithm performs relatively better for larger alphabets.
\item For a fixed alphabet size, there is a slight increase in the
  naive$/$KMP ratio, as the pattern length increases.
\item The distribution of the pattern occurrences in the text does not
  seem to make a change in performance.
\end{enumerate}
An analysis of these implementation behaviors appears in the next
subsection.

\subsubsection{Analysis}\label{sss:ext_anl}

We analyse all four results noted above.

{\bf Better Performance of the Naive Algorithm}\\
We have seen that the mean number of comparisons of the naive
algorithm for binary alphabets is bounded by
$$
n\sum_{i=1}^m  {i\over 2^i}\ {\rm which\ is\ bounded\ by}\ 2n\ {\rm
  comparisons}.
$$
The running time of the KMP algorithm is also bounded by
$O(2n)$. However, the control of the KMP algorithm is more complex
than that of the naive algorithm, which would indicate a constant
ratio in favor of the naive algorithm. However, when the KMP algorithm
encounters a mismatch it follows the failure link, which avoids the
need to re-check a larger substring. Thus, for longer length patterns,
where there are more possibilities of following the failure links for
longer distances, there is a lessening advantage of the naive
algorithm.

{\bf Better Performance of the Naive Algorithm for Larger Alphabets}\\
This is fairly clear when we realize that the mean performance of the
naive algorithm for alphabet of size $k$ is:
$$
n\sum_{i=1}^m  {i\over k^i}\ = n{k\over (k-1)^2}\ {\rm
  comparisons}.
$$
This is clearly decreasing the larger the alphabet size. However,
the repetitive traversal of the failure link, even in cases where
there is no equality in the comparison check, will still raise the
relative running time of the KMP algorithm. Here too, the longer the
pattern length, the more failure link traversals of the KMP, and thus
less overall comparisons, which slightly decreases the advantage of
the naive algorithm.

{\bf The Distribution of Pattern Occurrences in the Text}\\
If the pattern is not periodic, and if the patterns are not too
frequent in the text, then there will be at most one pattern in a text
substring of length $2m$. In these circumstances, there is really no
effect to the distribution of the pattern in the text. We would expect
a difference if the pattern is long with a small period. Indeed, an
extreme such case is tested in Subsection~\ref{sss:per}.

\subsubsection{A Very Structured Example}\label{sss:per}

\begin{figure}[htb]
\centering
\begin{minipage}{10cm}
  \includegraphics[width=\linewidth]{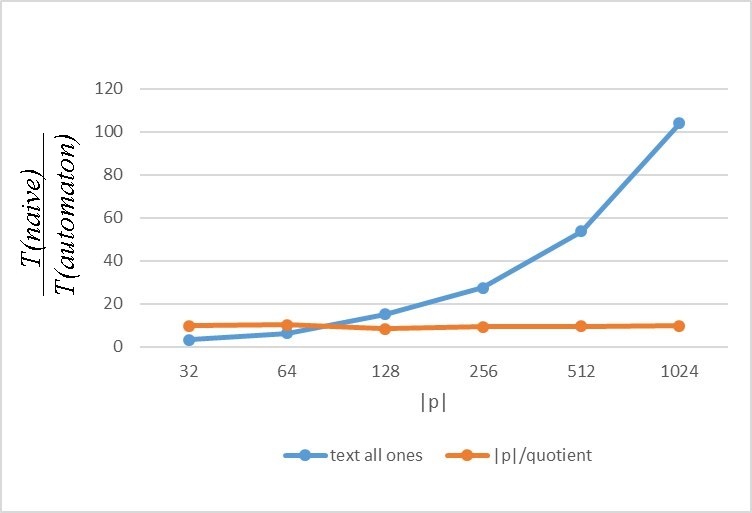}
  \caption{Performance in the Exact Matching case, periodic text.}
  \label{f:exactper}
\end{minipage}
\end{figure}

All previous analyses point to the conviction that the more times a
prefix of the pattern appears in the text, and the more periodic the
pattern, the better will be the performance of the KMP algorithm. The
most extreme case would be of text $A^n$ ($A$ concatenated $n$ times),
and pattern $A^{m-1}B$. Indeed the results of this case appear in
Fig.~\ref{f:exactper}.

Theoretical analysis of the naive algorithm predicts that we will have
$nm$ comparisons, where $n$ is the text length and $m$ is the
pattern length. The KMP algorithm will have $2n$ comparisons, for any
pattern length. Thus the ratio $q$ of naive to KMP will be $O({m\over
2})$. In fact, when we plot ${m\over q}$ we get twice the cost of the
control of the KMP algorithm. This can be seen in
Fig.~\ref{f:exactper} to be $5$.

\subsection{Parameterized Matching}\label{ss:par}

\subsubsection{Results} \label{sss:par_res}
The exact matching results behaved roughly in the manner we
expected. The surprise came in the parameterized matching case.
Below are the results of our tests.
As in the exact matching case, the tables show the alphabet size, the
pattern length, the average
of the running times of the naive algorithm for the 10 tests, the
average of the running time of the automaton-based algorithm for the
10 tests, and the ratio $q$ of the naive algorithm running time over the
automaton-based algorithm running time. Any ratio value {\em above}
$1$ means that the
{\em automaton-based algorithm is faster}. A {\em large} value indicates a
{\em better} performance of the automaton-based algorithm.

The following graphs show the results of our
tests for the different pattern lengths. The $x$-axis is the pattern
size. The $y$-axis is the ratio of the naive algorithm running time to
the automaton-based algorithm running time.  The different colors
depict alphabet sizes. To better see the effect of the pattern
distribution in the text, we also map, on the same graph, both
cases. In this graph, the $x$-axis is the average running time of {\em
  all} pattern lengths per alphabet size, and the $y$-axis is the
ratio of the naive algorithm
running time to the automaton-based algorithm running time. The
results of the uniformly random distribution are mapped in one color,
and the results of all pattern occurrences in the last half of the
text are mapped in another.

\begin{figure}[htb]
\centering
\begin{minipage}{10cm}
  \includegraphics[width=\linewidth]{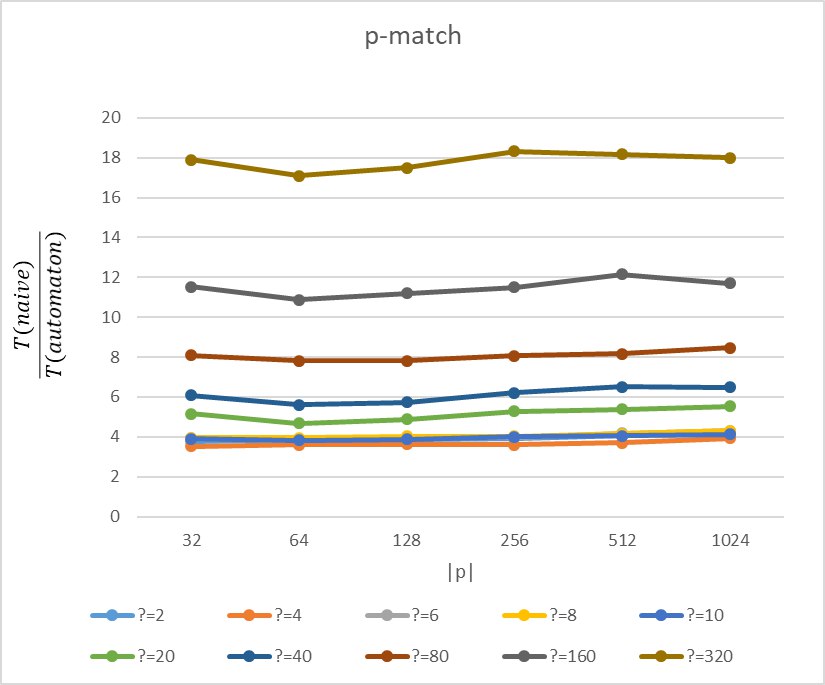}
  \caption{Performance in the Parameterized Matching case, pattern occurrences
    distributed uniformly random.}
  \label{f:paru}
\end{minipage}
\end{figure}

\begin{figure}[htb]
\centering
\begin{minipage}{10cm}
  \includegraphics[width=\linewidth]{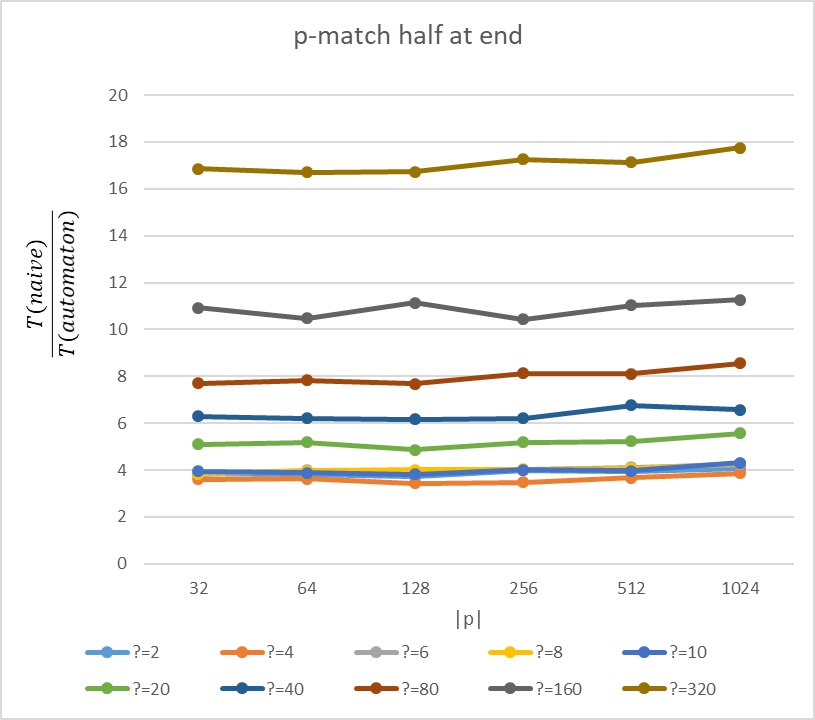}
  \caption{Performance in the Parameterized Matching case, pattern occurrences
  congregated at end of text.}
  \label{f:pare}
\end{minipage}
\end{figure}

\begin{figure}[htb]
\centering
\begin{minipage}{10cm}
  \includegraphics[width=\linewidth]{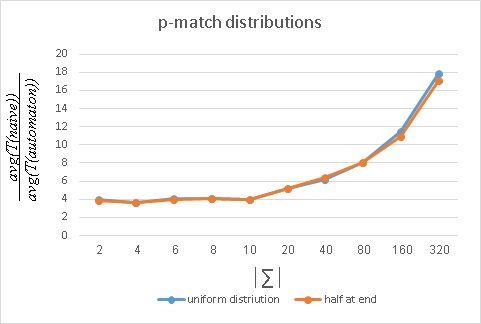}
  \caption{Comparison of average performance of uniform pattern
    distribution vs. pattern occurrences congregated at end of text.}
  \label{f:parb}
\end{minipage}
\end{figure}

The parameterized matching results appear in tables~\ref{t:ext_tbl3} and
~\ref{t:ext_tbl4} in the appendix. Figs.~\ref{f:paru} and~\ref{f:pare}
map the results of the parameterized matching comparisons for the case
where the  patterns were inserted at random vs. the case where the
patterns appear at the last half of the text. In Fig.~\ref{f:parb} we
map {\em at the same graph} the average results of both the cases where the
patterns appear at the text uniformly at random, and where the
patterns appear at the last half of the text.

The results are very different from the exact matching
case. We note the following phenomena:
\begin{enumerate}
\item The automaton-based algorithm {\bf always performs significantly
  better} than the
  naive algorithm.
\item The automaton-based algorithm performs relatively better for
  larger alphabets. 
\item For a fixed alphabet size, the pattern length does not seem to make
  much difference.
\item The distribution of the pattern occurrences in the text does not
  seem to make a change in performance.
\end{enumerate}
An analysis of these implementation behaviors and an explanation of
the seemingly opposite results from the exact matching case appear in
the next subsection.

\subsubsection{Analysis}\label{sss:par_anl}

We analyse all four results noted above.

{\bf Better Performance of the Automaton-based Algorithm}\\
We have established that the mean number of comparisons for the naive
algorithm in size $k$ alphabet is
$$
n\sum_{i=1}^m  {i\over k^i}\ = n{k\over (k-1)^2}\ {\rm
  comparisons}.
$$
However, when it comes to parameterized matching, any order of the
alphabet symbols is a match, thus the mean number of comparisons is to
be multiplied by $k!$. Therefore, for size $2$ alphabet we get $4n$
comparisons, and the number rises exponentially with the alphabet
size. Also, the automaton-based algorithms is constant at $2n$
comparisons. Even for a size $2$ alphabet, there is twice the number
of comparisons in the naive algorithm than in the automaton-based
algorithm. Note, also, that because of the need to find the last
parameterized match, the control mechanism even of the naive
algorithm, is more complex. This results in a superior performance of
the automaton-based algorithm even for small alphabets. Of course, the
larger the alphabet, the better the performance of the automaton-based
algorithm.

{\bf Pattern Length}

The pattern length does not play a role in the automaton-based
algorithm, where the number of comparisons is always bounded by $2n$. In
the naive case, the multiplication of the factorial of the alphabet
size is so overwhelming that it dominates the pattern length. For
example, note that for an extremely large alphabet, there would be a
leading prefix of different alphabet symbols. That prefix will always
be traversed by the naive algorithm. The larger the alphabet, the
longer will be the mean length of that prefix.

{\bf Pattern Distribution}

As in the exact matching case, for a non-periodic pattern that does
not appear too many times, the distribution of occurrences will have
no effect on the complexity.

\section{DNA Data}\label{s:dna}

Having understood the behavior of the naive algorithm and the
automaton-based algorithm in randomly generated texts, the natural
question is are there any ``real'' texts for which the naive algorithm
performs better than the automaton-based algorithm.

We performed the same experiments on DNA data. The experimental
setting was identical to that of the randomly generated texts with the
following differences:
\begin{enumerate}
\item The DNA of the fruit fly, {\em Drosophila melanogaster} is 143.7
  MB long. We extracted 60 subsequences of length 1,000,000
  each, as FASTA data from the NIH National Library of Medicine,
  National Center for Biotechnology Information. We ran 10 tests on
  each of the six pattern lengths 32, 64, 128, 256, 512, 1024. 
\item The alphabet size is 4, due to the four bases in DNA sequence.
\end{enumerate} 

Figs.~\ref{f:dnae} and ~\ref{f:dnap} below show the ratio between the
average running time of the naive algorithm and the automaton based
algorithm. As in the uniformly random text we see that for the exact
matching case the ratio is less than $1$, i.e., the naive algorithm is
faster, whereas in the parameterized matching case, the ratio is more
than 1, indicating that the automaton based algorithm is faster.

\begin{figure}[htb]
\centering
\begin{minipage}{10cm}
  \includegraphics[width=\linewidth]{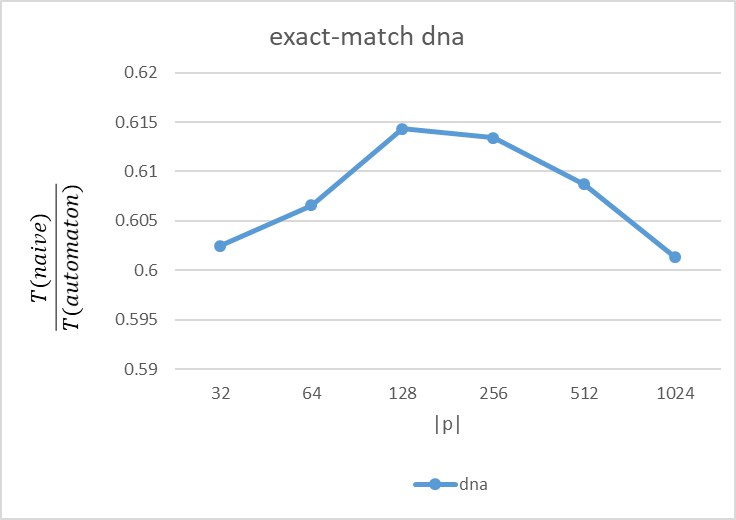}
  \caption{Performance in the Exact Matching case on DNA sequences.}
  \label{f:dnae}
\end{minipage}
\end{figure}

\begin{figure}[htb]
\centering
\begin{minipage}{10cm}
  \includegraphics[width=\linewidth]{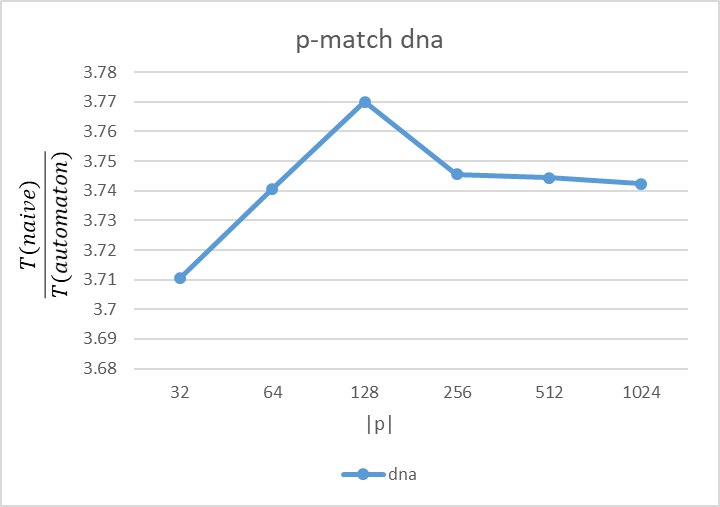}
  \caption{Performance in the Parameterized Matching case on DNA Sequences.}
  \label{f:dnap}
\end{minipage}
\end{figure}

\section{Conclusions}
The folk wisdom has always been that the naive string matching
algorithm will outperform the automaton-based algorithm for uniformly
random texts. Indeed this turns out to be the case for {\em exact
 matching}. This study shows that this is not the case for parameterized
matching, where the automaton-based algorithm {\em always outperforms}
the naive algorithm. This advantage is clear and is impressively
better the larger the alphabets. The same result is true for searches
over DNA data.

The conclusion to take away from this study is that one should not
automatically assume that the naive string matching algorithm is
better. The matching relation should be
analysed. There are various matchings for which an automaton-based
algorithm exists. We considered here parameterized matching, but other
matchings, such as ordered
matching~\cite{cnps:13,cikklprrw:13,kanps:14}, or Cartesian tree
matching~\cite{palp:19,pbalp:20,SGRFLP:21}, can
also be solved by automaton-based methods. In a practical application
it is worthwhile spending some time considering the type of matching
one is using. It may turn out to be that the automaton-based algorithm
will perform significantly better than the naive, even for uniformly
random texts. Alternately, even non-uniformly random  data may be such
that the naive algorithm performs better than the automaton based
algorithm for exact matching.

An open problem is to compare the search time in DNA
data to the search time in uniformly random data. While it is clear
that DNA data is not uniformly random, it would be interesting to
devise an experimental setting to compare search efficiency in both
types of strings.

\bibliographystyle{plain}
\bibliography{paper}

\newpage
\section{Appendix}\label{s:a}

\begin{table}[h!]
\centering
\begin{tabular}{|c|c|l|l|l|||c|c|l|l|l|}
\hline
{\bf $|\Sigma |$} &{\bf patt. length}  &{\bf Naive}  &{\bf KMP}  &
{\bf ${Naive\over KMP}$} &
{\bf $|\Sigma |$} &{\bf patt. length}  &{\bf Naive}  &{\bf KMP}  &
{\bf ${Naive\over KMP}$}
\\
\hline
\hline
\multirow{6}{*}{2}
& 32 & 4514.1 & 6712.5 & 0.6729
& \multirow{6}{*}{4}& 32 & 3174.2 & 5409.9 & 0.5879\\
& 64 & 4449.3 & 6727.8 & 0.6623 &
& 64 & 3167.8 & 5428.3 & 0.5818\\
& 128 & 4697.3 & 6764.3 & 0.693 &
& 128 & 3136.8 & 5293.0 & 0.5917\\
& 256 & 4522.9 & 6814.2 & 0.6666 &
& 256 & 3109.7 & 5228.2 & 0.5942\\
& 512 & 4764.7 & 6734.7 & 0.7051 &
& 512 & 3108.8 & 5110.5 & 0.608\\
& 1024 & 4521.4 & 6188.7 & 0.7304 &
& 1024 & 3141.1 & 4928.7 & 0.6368\\
\hline
\hline
\multirow{6}{*}{6}
& 32 & 2225.1 & 4331.2 & 0.5139
& \multirow{6}{*}{8}& 32 & 1771.8 & 3903.4 & 0.4553\\
& 64 & 2199.2 & 4263.2 & 0.5157 &
& 64 & 1794.5 & 3852.4 & 0.4659\\
& 128 & 2180.9 & 4270.6 & 0.5108 &
& 128 & 1764.0 & 3789.7 & 0.4654\\
& 256 & 2169.2 & 4201.4 & 0.5163 &
& 256 & 1766.5 & 3798.4 & 0.4652\\
& 512 & 2193.2 & 4128.4 & 0.5314 &
& 512 & 1771.9 & 3670.6 & 0.4827\\
& 1024 & 2238.7 & 4110.1 & 0.5455 &
& 1024 & 1827.3 & 3596.6 & 0.5085\\
\hline
\hline
\multirow{6}{*}{10}
& 32 & 1593.1 & 3598.9 & 0.4427
& \multirow{6}{*}{20} & 32 & 1312.0 & 3309.2 & 0.396\\
& 64 & 1578.3 & 3586.4 & 0.44 &
& 64 & 1428.6 & 3297.7 & 0.4269\\
& 128 & 1564.8 & 3563.9 & 0.4391 &
& 128 & 1252.7 & 3264.9 & 0.3817\\
& 256 & 1594.5 & 3531.6 & 0.4516 &
& 256 & 1187.4 & 3161.3 & 0.375\\
& 512 & 1554.3 & 3626.0 & 0.4317 &
& 512 & 1281.7 & 3166.8 & 0.4\\
& 1024 & 1892.5 & 3380.0 & 0.5619 &
& 1024 & 1274.6 & 2923.1 & 0.4347\\
\hline
\hline
\multirow{6}{*}{40}
& 32 & 943.9 & 2846.7 & 0.3316
& \multirow{6}{*}{80} & 32 & 898.1 & 2758.7 & 0.3242\\
& 64 & 964.3 & 2869.3 & 0.3358 &
& 64 & 938.4 & 2777.9 & 0.335\\
& 128 & 972.5 & 2852.5 & 0.3401 &
& 128 & 946.7 & 2824.5 & 0.3336\\
& 256 & 952.6 & 2835.3 & 0.3363 &
& 256 & 875.9 & 2709.0 & 0.323\\
& 512 & 975.4 & 2769.0 & 0.3523 &
& 512 & 875.8 & 2653.9 & 0.3302\\
& 1024 & 970.5 & 2655.4 & 0.3655 &
& 1024 & 899.6 & 2605.0 & 0.346\\
\hline
\hline
\multirow{6}{*}{160}
& 32 & 810.9 & 2686.1 & 0.302
& \multirow{6}{*}{320} & 32 & 790.3 & 2712.0 & 0.2916\\
& 64 & 794.0 & 2733.1 & 0.2918 &
& 64 & 833.4 & 2711.1 & 0.3074\\
& 128 & 922.2 & 2771.1 & 0.3281 &
& 128 & 803.3 & 2676.3 & 0.3005\\
& 256 & 899.2 & 2700.6 & 0.3285 &
& 256 & 785.2 & 2743.0 & 0.2877\\
& 512 & 897.8 & 2635.6 & 0.3374 &
& 512 & 878.5 & 2690.4 & 0.3269\\
& 1024 & 861.6 & 2534.9 & 0.3399 &
& 1024 & 883.8 & 2563.6 & 0.3427\\
\hline
\hline
\end{tabular}
\caption{Implementation Results - Exact Matching, patterns uniformly
  distributed.}
\label{t:ext_tbl1}
\end{table}

\newpage
\begin{table}[h!]
\centering
\begin{tabular}{|c|c|l|l|l|||c|c|l|l|l|}
\hline
{\bf $|\Sigma |$} &{\bf patt. length}  &{\bf Naive}  &{\bf KMP}  &
{\bf ${Naive\over KMP}$} &
{\bf $|\Sigma |$} &{\bf patt. length}  &{\bf Naive}  &{\bf KMP}  &
{\bf ${Naive\over KMP}$}
\\
\hline
\hline
\multirow{6}{*}{2}
& 32 & 4613.3 & 6931.1 & 0.6649
& \multirow{6}{*}{4}
& 32 & 3091.7 & 5362.9 & 0.5759\\
& 64 & 4570.1 & 6695.7 & 0.6824 &
& 64 & 3203.2 & 5499.5 & 0.5819\\
& 128 & 4462.8 & 6702.2 & 0.667 &
& 128 & 3190.4 & 5373.6 & 0.5933\\
& 256 & 4441.5 & 6644.9 & 0.6692 &
& 256 & 3200.3 & 5413.1 & 0.5924\\
& 512 & 4786.4 & 6441.1 & 0.744 &
& 512 & 3305.2 & 5340.0 & 0.6176\\
& 1024 & 4493.8 & 6360.6 & 0.7105 &
& 1024 & 3322.4 & 5125.8 & 0.6469\\
\hline
\hline
\multirow{6}{*}{6}
& 32 & 2374.7 & 4638.6 & 0.509
& \multirow{6}{*}{8}
& 32 & 1836.3 & 3978.1 & 0.4616\\
& 64 & 2336.6 & 4586.8 & 0.5093 &
& 64 & 1804.2 & 3930.2 & 0.4589\\
& 128 & 2467.1 & 4597.0 & 0.534 &
& 128 & 1816.9 & 3908.6 & 0.465\\
& 256 & 2350.4 & 4453.1 & 0.5274 &
& 256 & 1802.8 & 3875.2 & 0.4655\\
& 512 & 2306.2 & 4447.2 & 0.5243 &
& 512 & 1792.0 & 3832.8 & 0.4684\\
& 1024 & 2411.2 & 4302.9 & 0.5597 &
& 1024 & 1889.1 & 3640.7 & 0.5183\\
\hline
\hline
\multirow{6}{*}{10}
& 32 & 1741.8 & 3762.0 & 0.4608
& \multirow{6}{*}{20}
& 32 & 1242.4 & 3173.7 & 0.3916\\
& 64 & 1719.8 & 3772.8 & 0.4528 &
& 64 & 1173.5 & 3251.9 & 0.3615\\
& 128 & 1616.5 & 3800.2 & 0.4264 &
& 128 & 1286.4 & 3302.4 & 0.3847\\
& 256 & 1685.1 & 3814.7 & 0.4424 &
& 256 & 1334.3 & 3234.5 & 0.411\\
& 512 & 1774.0 & 3724.7 & 0.4737 &
& 512 & 1231.7 & 3090.4 & 0.399\\
& 1024 & 1727.8 & 3484.3 & 0.4922 &
& 1024 & 1263.8 & 3031.5 & 0.4168\\
\hline
\hline
\multirow{6}{*}{40}
& 32 & 1108.6 & 3048.3 & 0.3606
& \multirow{6}{*}{80}
& 32 & 867.4 & 2912.6 & 0.2988\\
& 64 & 1014.5 & 3084.3 & 0.3283 &
& 64 & 941.2 & 2912.8 & 0.3248\\
& 128 & 1142.9 & 3210.4 & 0.3533 &
& 128 & 1023.5 & 2872.7 & 0.3546\\
& 256 & 1026.3 & 3005.2 & 0.3413 &
& 256 & 1005.4 & 2949.3 & 0.3397\\
& 512 & 1503.7 & 2930.9 & 0.5205 &
& 512 & 956.0 & 2852.1 & 0.3355\\
& 1024 & 1170.1 & 2926.9 & 0.3951 &
& 1024 & 954.3 & 2701.8 & 0.3532\\
\hline
\hline
\multirow{6}{*}{160}
& 32 & 981.8 & 2855.0 & 0.3393
& \multirow{6}{*}{320}
& 32 & 769.6 & 2662.8 & 0.2894\\
& 64 & 863.6 & 2818.4 & 0.3061 &
& 64 & 771.8 & 2681.5 & 0.2882\\
& 128 & 908.6 & 2842.8 & 0.3178 &
& 128 & 799.5 & 2627.0 & 0.304\\
& 256 & 851.2 & 2796.4 & 0.3047 &
& 256 & 917.9 & 2722.0 & 0.3345\\
& 512 & 909.6 & 2917.1 & 0.313 &
& 512 & 967.3 & 2757.1 & 0.3455\\
& 1024 & 1174.9 & 2815.9 & 0.4093 &
& 1024 & 951.2 & 2601.3 & 0.3604\\
\hline
\hline
\end{tabular}
\caption{Implementation Results - Exact Matching, patterns at end.}
\label{t:ext_tbl2}
\end{table}

\newpage
\begin{table}[h!]
\centering
\begin{tabular}{|c|c|l|l|l|||c|c|l|l|l|}
\hline
{\bf $|\Sigma |$} &{\bf patt. length}  &{\bf Naive}  &{\bf KMP}  &
{\bf ${Naive\over KMP}$} &
{\bf $|\Sigma |$} &{\bf patt. length}  &{\bf Naive}  &{\bf KMP}  &
{\bf ${Naive\over KMP}$}
\\
\hline
\hline
\multirow{6}{*}{2}
& 32 & 25738.0 & 6871.8 & 3.7655
& \multirow{6}{*}{4}
& 32 & 26104.6 & 7489.6 & 3.5351\\
& 64 & 25996.5 & 6761.4 & 3.8593 &
& 64 & 26734.4 & 7538.6 & 3.5998\\
& 128 & 26080.5 & 6780.8 & 3.8571 &
& 128 & 26281.4 & 7370.8 & 3.6136\\
& 256 & 26269.7 & 6688.6 & 3.934 &
& 256 & 26204.3 & 7361.0 & 3.6062\\
& 512 & 26004.0 & 6440.3 & 4.0456 &
& 512 & 26169.6 & 7123.6 & 3.71\\
& 1024 & 26456.0 & 6277.9 & 4.2167 &
& 1024 & 26570.9 & 6863.1 & 3.924\\
\hline
\hline
\multirow{6}{*}{6}
& 32 & 26213.2 & 6818.3 & 3.96
& \multirow{6}{*}{8}
& 32 & 26863.5 & 7229.3 & 3.9411\\
& 64 & 26244.3 & 7022.8 & 3.8621 &
& 64 & 27010.3 & 7258.5 & 3.9394\\
& 128 & 26130.3 & 6879.7 & 3.9429 &
& 128 & 26965.3 & 7067.4 & 4.0336\\
& 256 & 26141.2 & 6778.1 & 3.987 &
& 256 & 26918.8 & 7099.7 & 4.0304\\
& 512 & 26212.3 & 6460.7 & 4.1752 &
& 512 & 27211.8 & 6888.9 & 4.1592\\
& 1024 & 26171.5 & 6312.7 & 4.2986 &
& 1024 & 27406.5 & 6698.6 & 4.3042\\
\hline
\hline
\multirow{6}{*}{10}
& 32 & 28663.6 & 7629.8 & 3.8967
& \multirow{6}{*}{20}
& 32 & 28539.6 & 5832.4 & 5.1463\\
& 64 & 28787.8 & 7787.6 & 3.8351 &
& 64 & 28543.3 & 6329.9 & 4.6772\\
& 128 & 28629.8 & 7664.8 & 3.8775 &
& 128 & 28254.3 & 6041.4 & 4.8694\\
& 256 & 28647.0 & 7478.5 & 3.99 &
& 256 & 28526.7 & 5733.2 & 5.2725\\
& 512 & 28843.4 & 7406.5 & 4.0576 &
& 512 & 28326.8 & 5546.4 & 5.3728\\
& 1024 & 28516.9 & 7074.3 & 4.1282 &
& 1024 & 28457.7 & 5433.1 & 5.5292\\
\hline
\hline
\multirow{6}{*}{40}
& 32 & 33994.8 & 5708.6 & 6.0731
& \multirow{6}{*}{80}
& 32 & 42524.1 & 5292.8 & 8.0792\\
& 64 & 33826.0 & 6076.9 & 5.6046 &
& 64 & 41425.9 & 5340.1 & 7.8236\\
& 128 & 33971.3 & 5994.7 & 5.7342 &
& 128 & 41547.1 & 5387.7 & 7.8057\\
& 256 & 33740.9 & 5544.9 & 6.2016 &
& 256 & 41489.1 & 5269.7 & 8.0644\\
& 512 & 34501.6 & 5411.8 & 6.5045 &
& 512 & 41615.2 & 5189.5 & 8.165\\
& 1024 & 34172.0 & 5353.9 & 6.496 &
& 1024 & 42184.8 & 5067.8 & 8.478\\
\hline
\hline
\multirow{6}{*}{160}
& 32 & 54881.0 & 4789.375 & 11.5167
& \multirow{6}{*}{320}
& 32 & 70360.0 & 3919.7 & 17.9046\\
& 64 & 56750.0 & 5222.7 & 10.8806 &
& 64 & 75533.8 & 4456.5 & 17.1093\\
& 128 & 57775.6 & 5212.2 & 11.2048 &
& 128 & 75098.4 & 4284.8 & 17.4987\\
& 256 & 56719.3 & 4953.4 & 11.5 &
& 256 & 77763.7 & 4238.4 & 18.328\\
& 512 & 58276.6 & 4793.2 & 12.1498 &
& 512 & 75922.3 & 4181.3 & 18.1751\\
& 1024 & 57331.2 & 4913.2 & 11.7029 &
& 1024 & 76831.3 & 4366.4 & 17.989\\
\hline
\hline
\end{tabular}
\caption{Implementation Results - Parameterized Matching, patterns
  uniformly distributed.}
\label{t:ext_tbl3}
\end{table}

\newpage
\begin{table}[h!]
\centering
\begin{tabular}{|c|c|l|l|l|||c|c|l|l|l|}
\hline
{\bf $|\Sigma |$} &{\bf patt. length}  &{\bf Naive}  &{\bf KMP}  &
{\bf ${Naive\over KMP}$} &
{\bf $|\Sigma |$} &{\bf patt. length}  &{\bf Naive}  &{\bf KMP}  &
{\bf ${Naive\over KMP}$}
\\
\hline
\hline
\multirow{6}{*}{2}
& 32 & 26063.4 & 6801.4 & 3.8439
& \multirow{6}{*}{4}
& 32 & 26616.5 & 7505.3 & 3.61\\
& 64 & 26285.3 & 6878.0 & 3.828 &
& 64 & 26571.7 & 7443.4 & 3.6226\\
& 128 & 26053.8 & 7047.4 & 3.7047 &
& 128 & 26385.6 & 7829.9 & 3.4449\\
& 256 & 26612.5 & 6671.7 & 3.996 &
& 256 & 26236.1 & 7649.5 & 3.4807\\
& 512 & 26501.7 & 6764.8 & 3.9329 &
& 512 & 26660.5 & 7356.9 & 3.6748\\
& 1024 & 26397.8 & 6506.4 & 4.0685 &
& 1024 & 26667.6 & 7038.6 & 3.8591\\
\hline
\hline
\multirow{6}{*}{6}
& 32 & 26312.4 & 7071.4 & 3.828
& \multirow{6}{*}{8}
& 32 & 27246.5 & 7421.6 & 3.8491\\
& 64 & 26733.2 & 6924.6 & 3.9976 &
& 64 & 27046.1 & 7185.0 & 3.9748\\
& 128 & 26470.1 & 7067.1 & 3.8636 &
& 128 & 27117.6 & 7170.1 & 4.009\\
& 256 & 26346.3 & 6701.1 & 4.0218 &
& 256 & 27154.8 & 7089.7 & 4.04\\
& 512 & 26610.6 & 6682.2 & 4.117 &
& 512 & 26901.8 & 6998.0 & 4.0791\\
& 1024 & 26632.3 & 6399.8 & 4.2563 &
& 1024 & 27227.5 & 6667.8 & 4.2963\\
\hline
\hline
\multirow{6}{*}{10}
& 32 & 29612.8 & 7759.5 & 3.9578
& \multirow{6}{*}{20}
& 32 & 29588.6 & 6153.9 & 5.0995\\
& 64 & 28948.9 & 7748.2 & 3.8873 &
& 64 & 29393.4 & 6010.3 & 5.1754\\
& 128 & 29305.1 & 7925.5 & 3.829 &
& 128 & 29498.7 & 6312.8 & 4.8688\\
& 256 & 29457.3 & 7619.7 & 4.0189 &
& 256 & 29659.5 & 5966.3 & 5.1945\\
& 512 & 29650.7 & 7836.9 & 3.9754 &
& 512 & 29067.8 & 5802.3 & 5.226\\
& 1024 & 30742.0 & 7421.5 & 4.3099 &
& 1024 & 28922.4 & 5455.1 & 5.5624\\
\hline
\hline
\multirow{6}{*}{40}
& 32 & 34179.7 & 5577.9 & 6.2968
& \multirow{6}{*}{80}
& 32 & 41534.5 & 5441.2 & 7.6963\\
& 64 & 34385.3 & 5723.2 & 6.2199 &
& 64 & 41907.7 & 5373.0 & 7.8299\\
& 128 & 34951.8 & 5758.1 & 6.1685 &
& 128 & 41709.3 & 5474.4 & 7.6894\\
& 256 & 36703.8 & 6033.8 & 6.216 &
& 256 & 41900.3 & 5211.0 & 8.1372\\
& 512 & 37417.4 & 5682.4 & 6.7656 &
& 512 & 41753.4 & 5196.7 & 8.1023\\
& 1024 & 35190.1 & 5488.2 & 6.5708 &
& 1024 & 43312.9 & 5074.3 & 8.5567\\
\hline
\hline
\multirow{6}{*}{160}
& 32 & 52173.125 & 4773.875 & 10.9192
& \multirow{6}{*}{320}
& 32 & 67440.5 & 3981.8 & 16.8561\\
& 64 & 54173.0 & 5176.9 & 10.4658 &
& 64 & 71874.8 & 4294.4 & 16.7108\\
& 128 & 56313.9 & 5032.4 & 11.1442 &
& 128 & 72359.4 & 4315.2 & 16.725\\
& 256 & 54897.4 & 5257.0 & 10.433 &
& 256 & 72268.3 & 4179.1 & 17.2654\\
& 512 & 55123.0 & 5025.2 & 11.0268 &
& 512 & 72729.5 & 4234.7 & 17.1366\\
& 1024 & 55603.0 & 4915.0 & 11.26 &
& 1024 & 73777.8 & 4152.8 & 17.7372\\
\hline
\hline
\end{tabular}
\caption{Implementation Results - Parameterized Matching, patterns at end.}
\label{t:ext_tbl4}
\end{table}

\end{document}